\documentclass{aa}                             

\usepackage{graphicx}
\usepackage{array}
\usepackage{lscape}                                
\usepackage{txfonts}
\usepackage{natbib}
\usepackage{longtable}
\usepackage{color}
\usepackage{mathrsfs} 

\begin{document}


%

\title{The IACOB project\thanks{Based on observations made with the Nordic Optical Telescope, operated by NOTSA and the Mercator Telescope, operated by the Flemish Community, both at the Observatorio del Roque de los Muchachos (La Palma, Spain) of the Instituto de Astrof\'isica de Canarias.}}

\subtitle{IV. New predictions for high-degree non-radial mode instability domains in massive stars and connection with macroturbulent broadening}

\author{M.~Godart\inst{1,2,3}, S.~Sim\'on-D\'{\i}az\inst{1,2},  A.~Herrero\inst{1,2}, M.A.~Dupret\inst{3}, A.~Gr\"{o}tsch-Noels\inst{3}, S.J.A.J.~Salmon\inst{3,4},  P. Ventura\inst{5}}

\institute{Instituto de Astrof\'isica de Canarias, E-38200 La Laguna, Tenerife, Spain              
             \and
             Departamento de Astrof\'isica, Universidad de La Laguna, E-38205 La Laguna, Tenerife, Spain
             \and
             Institut d'Astrophysique et de G\'eophysique, Universit\`e de Li\`ege, 17 all\'ee du 6 Ao\^ut, 4000, Li\`ege, Belgium
             \and
             Laboratoire AIM, CEA/DSM-CNRS-Universit\'e Paris 7, Irfu/Service d'Astrophysique, CEA-Saclay, 91191 Gif-sur-Yvette, France
             \and
             INAF - Osservatorio Astronomico di Roma, Via Frascati 33, I-00040 Monte Porzio Catone (RM), Italy
             }
           
\offprints{mgodart@ulg.ac.be}

\date{Accepted}

\titlerunning{High-degree instability domains and connection to macroturbulent broadening in OB stars}

\authorrunning{Godart et al.}

%
\abstract{
Asteroseismology is a powerful tool to access the internal structure of stars. Apart from the important impact of theoretical developments, progress in this field has been commonly associated with the analysis of time-resolved observations. Recently, the so-called macroturbulent broadening has been proposed to be a complementary and less expensive way -- in terms of observational time -- to investigate pulsations in massive stars.}
{We assess to what extent this  ubiquitous non-rotational broadening component shaping the line profiles of O stars and B supergiants is a spectroscopic signature of pulsation modes driven by a heat mechanism.}
{We compute stellar main sequence and post-main sequence models from 3 to 70~${\rm M}_{\odot}$ with the ATON stellar evolution code and determine the instability domains for heat-driven modes for degrees $\ell$\,=\,1\,--\,20 using the adiabatic and non-adiabatic codes LOSC and MAD. We use the observational material presented in Sim\'on-D\'iaz et al. (2016) to investigate possible correlations between the single snapshot line-broadening properties of a sample of $\approx$~260 O and B-type stars and their location inside/outside the various predicted instability domains.}
{
We present an homogeneous prediction for the non-radial instability domains of massive stars for degree $\ell$ up to 20. 
We provide a global picture of what to expect from an observational point of view in terms of frequency range of excited modes, and investigate the behavior of the instabilities with stellar evolution and increasing degree of the mode. 
Furthermore, our pulsational stability analysis, once compared to the empirical results of Sim\'on-D\'iaz et al. (2016), indicates that stellar oscillations originated by a heat mechanism can not explain alone the occurrence of the large non-rotational line-broadening component commonly detected in the O star and B supergiant domain.} 
{}
%
\maketitle

\section{Introduction}\label{sec_intro}
In 1992, the new stellar opacities from \citeauthor*{Rogers1992} (computed with the OPAL code) revolutionized asteroseismology of massive stars by providing an explanation to what was called the $\beta$ Cephei problem: none of the commonly considered excitation mechanisms could account for the presence of pulsations in these B-type stars. As a consequence of these new computations, the mean Rosseland opacity was increased by a factor 3 at $\log T\simeq 5.2$, allowing the excitation of modes in these stars through the $\kappa$-mechanism in the iron opacity bump \citep[e.g.][]{Kiriakidis1992}. The inclusion of some additional metals in the OPAL computations by \cite{Iglesias1996} increased even further the opacity in that region. Meanwhile, similar results -- though with a slightly hotter metal opacity bump -- were also obtained in parallel by the OP project \citep{Badnell2005}.

\citet{Pam99} was the first to present an updated overview of the theoretical predictions for instability domains in the upper main sequence (MS) accounting for these new opacities. These domains, which mainly refer to the regions of the Hertzsprung-Russell (HR) diagram where the so-called $\beta$ Cephei and SPB (slowly pulsating B) stars are located,  were subsequently revised by \citet{Miglio2007a,Miglio2007b} considering the updated solar metal mixture\footnote{Previous computations used the solar metal mixture proposed by \citet[][]{Grevesse1993}.} of \citet[][]{Asplund2005}, and different choices in the assumed opacities \cite[see also][]{Pamyatnykh2007}.

All these computations improved the agreement between models and observations.   However, some problems still remained. For example, standard theoretical models were not able to fit all the observed frequencies in some Galactic $\beta$ Cephei stars \citep[e.g. $12$~Lac and $\nu$~Eri,][]{Ausseloos2004,Pamyatnykh2004,Dziembowski2008,Desmet2009}. In addition, some Galactic late O-type stars were found to exhibit $\beta$~Cephei-type pulsations, while they are located outside the predicted instability domains \citep[e.g.,][]{Briquet2011}.  Moreover, with the progress in detection capabilities, B pulsating stars were also discovered in the Small Magellanic Cloud \citep[e.g.][and references therein]{Kou14} while, adopting the typical metallicity of this galaxy 
\citep[$Z\approx 0.003$, e.g.][]{Buchler2008}, 
no modes could be excited in a theoretical standard $\beta$ Cephei model \citep{Salmon2009}. The solution to these discrepancies between theoretical predictions and observations may still be related to the considered opacities \citep[see, e.g.,][]{Salmon2012,TurckChieze2013,Mor16}, or to the existence of other type of driving mechanisms of stellar oscillations not accounted for in the computations.

Moving up in the HR diagram to the O star and B supergiant domain (hereafter OB stars), we enter in a more uncharted territory in terms of instability predictions, in particular for the more evolved models. Due to the very high contrast in density between the core and the superficial layers in post-MS models (compared to MS models for the same initial mass), the number of nodes of the pulsation modes in the central layers become extremely high and numerical problems appear. But still it has become clear from both an observational and a theoretical point of view that OB stars present various types of oscillations. 
Indeed, as an extension to the $\beta$ Cephei and SPB instability strips in the B star domain, these more massive stars are predicted to present pressure (p-) and gravity (g-) pulsation modes excited by the $\kappa$-mechanism.
Depending on the stellar effective temperature ($T_{\rm eff}$), the iron or the helium opacity bumps are the main responsible for the occurrence of these modes in stars with masses $\gtrsim 9\, {\rm M}_{\odot}$ \citep[see, e.g.,][]{Dziembowski1993, Pam99, Saio2011}.
In addition, \cite{Saio2011} showed that oscillatory convective modes are expected to be excited in the HR diagram in a large portion of the MS and post-MS region of the massive star domain. These are modes punctually excited thanks to the convective motions in the iron and helium opacity bumps. Furthermore, modes excited by the $\epsilon$-mechanism \citep[e.g.][]{Noe86, Scu86, Unn89}, strange mode instabilities \citep[e.g.][and references therein]{Saio1998,Gla09}, and even stochastically excited waves excited by various driving mechanisms \citep[][]{Belkacem2010,Samadi2010,Shi13,Mat14,Aer15,Gra15} are predicted in massive OB stars. Actually, some of these predictions have already found empirical support \citep[e.g.,][]{Aer10,Degroote2010,Mor12,Buy15}. Recent reviews describing the state-of-the-art of the theoretical knowledge about pulsations in massive OB stars can be found in \cite{Saio2011,Godart2014,Saio2015} and \cite{Samadi2015}.

The slow but sure increase in the amount of observational data coming from space missions (e.g. MOST, CoRoT, Kepler, K2, BRITE), supported by on-ground spectroscopic material is not only showing us the rich and complex variety of pulsational-type phenomena present in the OB star domain \citep[see][and references therein]{Aer15a}, but is also providing us invaluable empirical information to step forward in the characterization and understanding of the pulsational properties of these massive stellar objects. This situation will improve thanks to future missions as TESS and PLATO, or coordinated on-ground telescope facility networks (e.g. SONG). 
It is timely to combine all the promptly available empirical information with our current theoretical knowledge of stellar oscillations in massive stars. In this regards, it is interesting to note that despite the remarkable progress made in the last decades we still lack from a comprehensive, homogeneous, in-depth pulsational stability analysis in the massive star domain (from the zero age MS to the more evolved phases and the complete massive star range above 3~${\rm M}_{\odot}$). Indeed, results are limited to low degree modes ($\ell\le3$) even combining heterogeneous computations by several authors \citep[][]{Miglio2007a,Pamyatnykh2007,Godart2011,Saio2011,Das13,Saio2015,Mor16}, except for the study of \citet{Bal99} that limited their high-degree computations to MS stars.

Higher-degree modes are usually disregarded because of the difficulty to be detected photometrically \citep[e.g.][]{Bal99,Aer10b} but can be important when interpreting spectroscopic observations \cite[see, e.g.,][]{Sch97}. An example of this situation was highlighted in \cite{Sim16}, where we presented the first comprehensive attempt to evaluate empirically the proposed hypothesis that the so-called macroturbulent broadening in O and B stars is produced by the collective effect of multiple non-radial pulsation modes \citep[e.g.][]{Luc76,Aer09}. Using high-spectral resolution, single snapshot observations of $\approx$~430 Galactic stars\footnote{From the IACOB spectroscopic database \citep[last described in][]{Sim15b}.} with spectral types ranging from O4 to B9, we investigated the potential connection between stellar oscillations driven by a heat mechanism and this ubiquitous and dominant non-rotational broadening component shaping the line profiles of O stars and B Supergiants. The result was not very promising when accounting only for dipole modes predictions ($\ell$=1). However, the consideration of a larger number of modes of higher degree might help improving the situation.
As a continuation to the work by \cite{Godart2011} and motivated by these recent advances in the investigation of the macroturbulent broadening in the OB star domain, 
we present in this paper new theoretical predictions for instability domains of $\ell\le 20$ non-radial modes for stars with masses between 3 and 70~${\rm M}_{\odot}$. We aim at providing a global homogeneous overview of the instability domains predicted in the upper part of the HR diagram together with some indications about the characteristics (in terms of type of modes and frequency spectrum) of the various unstable modes found in the whole region. We also expect that this theoretical work will become a useful tool for the interpretation of data gathered in on-going and future observational campaigns, especially in spectroscopy.

The paper is structured as follows. The model input physics is presented in Sect.~\ref{sec_computations}, while we introduce in Sect.~\ref{sec_diagnostic} the diagnostic diagrams used for the determination of the instability boundaries for pulsations. The instability domains resulting from a comprehensive pulsational instability analysis for MS and post-MS models from 3 to 70~${\rm M}_{\odot}$ for $\ell$=1 and 2 $\le \ell \le$ 20 are presented in Sects.~\ref{sub_bi_l_equal1} and \ref{sub_bi_l_larger1}, respectively. This is complemented with a discussion about the behavior
of high-degree modes as the stars evolve from the MS to the post-MS in Sect.~\ref{sub_global_l}. We then investigate in Sect.~\ref{sec_macro} possible correlations between the location of stars characterized by having an important contribution of so-called macroturbulent broadening to their line profiles and the predicted instability domains associated with heat-driven pulsating modes (Sect.~\ref{sec_macro_shr}). We then discuss the physical ingredients of the models that can alter the results, such as a change in the metallicity or in the adopted opacities (Sect.~\ref{sec_zams}). With this study we want to assess whether this spectroscopic feature is an observational signature of the existence of a distribution of motions in the line-formation region induced by this specific type of stellar pulsations. Last, other possible sources of macroturbulent broadening are briefly discussed in Sect.~\ref{sec_sources}, and the main conclusions of our work -- along with some future prospects -- are summarized in Sect.~\ref{sec_conclusions}.

\section{Stellar models and non-radial modes}\label{sec_computations}

We concentrate in this paper on the instability domain predictions for heat-driven pulsation modes produced by the $\kappa$-mechanism in the iron opacity bump. 
To this aim, we benefit from the computations performed in \citet{Godart2011}. The stellar models were computed using the ATON evolutionary code \citep{Ventura2008}. In order to cover the whole mass range of observations, we extended the grid of stellar models to 3\,--\,70~${\rm M}_{\odot}$. 
 We briefly mention here the main physical ingredients of these models: 
 \begin{itemize}
  \item convective transport is treated following the Mixing Length Theory of convection \citep{Bohm-Vitense1958} and by adopting the Schwarzschild criterion for convection \citep{Schwarzschild1906}, 
  \item no overshooting is included: asteroseismology of massive stars shows a rather large dispersion in the determination of the overshooting parameter (e.g. \citeauthor[][]{Noe15} \citeyear[][]{Noe15}, though \citeauthor{Cas14} \citeyear{Cas14} have argued that a mass dependent overshooting is necessary to recover the HR diagram features at high masses), 
  \item mass loss is taken into account for masses larger than 7~${\rm M}_{\odot}$ following the prescription of \cite{Vink2001}, 
  \item we used the OPAL opacity tables \citep{Iglesias1996}, extended  with the opacities of \cite{Ferguson2005} for $T<6000$~K,
  \item the considered metallicity is $Z=0.015$, and we used the metal mixture determined by \cite{Grevesse1993},
  \item microscopic diffusion and radiative levitation are ignored.
 \end{itemize}
 
At this point, we remark that our main set of computations (see also below) do not use the most recently accepted values of metallicity \citep[$Z=0.014$ and the metal mixture from][]{Asplund2009}. In addition, while we considered the OPAL opacity tables, we are aware of the availability of other computations, such as the  OP tables which favor the excitation of modes in the considered domain of stellar parameters \citep[e.g.][]{Miglio2007a,Pamyatnykh2007}. The computations performed here are still valid for the main purpose of this work, namely the investigation of the possible connection between heat driven modes and the so-called macroturbulent broadening in O and B stars. A detailed explanation of the reason of using the set of assumption quoted above, along with notes about the effect of considering a higher metallicity, a different metal mixture and/or the opacities from the OP project on the computed instability domains and, therefore, on the conclusions of our study is presented in Sect.~\ref{sec_zams}.

Adiabatic and non-adiabatic computations for non-radial pulsations in the whole range of masses described above were performed using the Li{\`e}ge Oscillation code  \citep[LOSC,][]{Scu08a} and the non-adiabatic code MAD \citep{Dupret2003} combined with the numerical method described in \citet{Godart2009} for the computations of  g-modes in models presenting an intermediate convective shell surrounding the He core. This method allows us to avoid the computation of the full dense spectrum of non-radial modes by pre-selecting the modes that are reflected at the bottom of the convective shell and potentially excited. 
The adiabatic and non-adiabatic frequencies were computed in the range $\omega=$0.05 to 20 where $\omega$ is the dimensionless frequency\footnote{The dimensionless frequency is defined as $\omega = \, \sigma \, \tau_{\sf dyn} \,= \, 2\,\pi\, f \, \tau_{\sf dyn}$ where $f$ is the frequency in Hz and $\tau_{\sf dyn}=\sqrt{R^3/GM}$ is the dynamical time in seconds. For the fundamental radial mode, $\omega$ is of the order of 3.}. This range covers periods from minutes to several hundreds of days.  
The degree $\ell$ of the mode was chosen to vary from 1 to 20 considering zonal modes only ($m=0$).
We remark the novelty and importance of the $\ell>$~3 (and up to 20) computations for the purpose of the study presented here. 
As already mentioned, higher-degree modes are usually disregarded because of the difficulty to be detected using photometric observations \citep[the amplitudes are averaged out over the unresolved stellar disc, see][for an analysis of the visibility of high-degree modes]{Bal99}. However, these modes, especially when they are combined together in a dense frequency spectrum, have a more important effect on the line profiles. Therefore, they become interesting in a spectroscopic context and, in particular, for the investigation of line-profile variability and line-broadening effects \citep[see, e.g.,][]{Sch97, Aer09, Sim16}. 

To identify the excited modes, we first evaluate whether the work produced by the star during a pulsation cycle is positive \citep[e.g.][]{Unn89}.
This criterion, used in all non-adiabatic codes, gives 
a first indication on the stability of the pulsation. However, it may not be sufficiently reliable when studying instabilities of massive stars. Under certain circumstances, some modes, which would be considered excited according to the work integral criterion, present peculiar eigenfunctions. In particular, in the most massive and evolved models, we have found excited modes that are misinterpreted by the non-adiabatic code due to the huge number of nodes compared to the limited number of mesh points we imposed in the code. Indeed, the asymptotic formula for g-modes \citep[][]{Tas80}
\begin{equation}
k^2\approx \frac{N^2}{\sigma^2}\frac{\ell(\ell+1)}{r^2} \,\,\, \rm{( if}\,\, \sigma\ll N,L_\ell\,\rm{)}
\end{equation}
shows that for a given frequency ($\sigma$), the number of nodes ($k$) for a mode is very large in the region of the peaks of $N^2$ (i.e., the Brunt-V\"{a}is\"{a}l\"{a} frequency, see Eq.~\ref{eq-bv}) and/or for large values of $\ell$. When studying the eigenfunctions in detail, these modes are actually not physically excited. 

These issues are recurrent over the whole spectra of computed frequencies for the most evolved and/or the most massive models, but the modes with the largest periods are the most problematic to compute due to their very large number of nodes (for $\omega\lesssim 0.2$ though this upper limit increases with increasing $\ell$).
Therefore, we  decided to inspect visually every modal diagram (see Sect.~\ref{sec_diagnostic}) resulting from the non-adiabatic code computations to identify the potential unphysical modes. For all these dubious modes, we evaluate the reliability of the results by studying in detail the corresponding eigenfunctions. In particular, whenever necessary, we increased the number of mesh points (usually around 5000 points) in the most critical regions of the star such as, for example, in the non-zero mean molecular weight gradient zone or where the work produced by the star reaches high absolute values. As a result, we ended up with what we call {\em cleaned} modal diagrams in which only these physically meaningful modes are included.

\begin{figure*}[!t]
\centering
\includegraphics[width=0.75\textwidth]{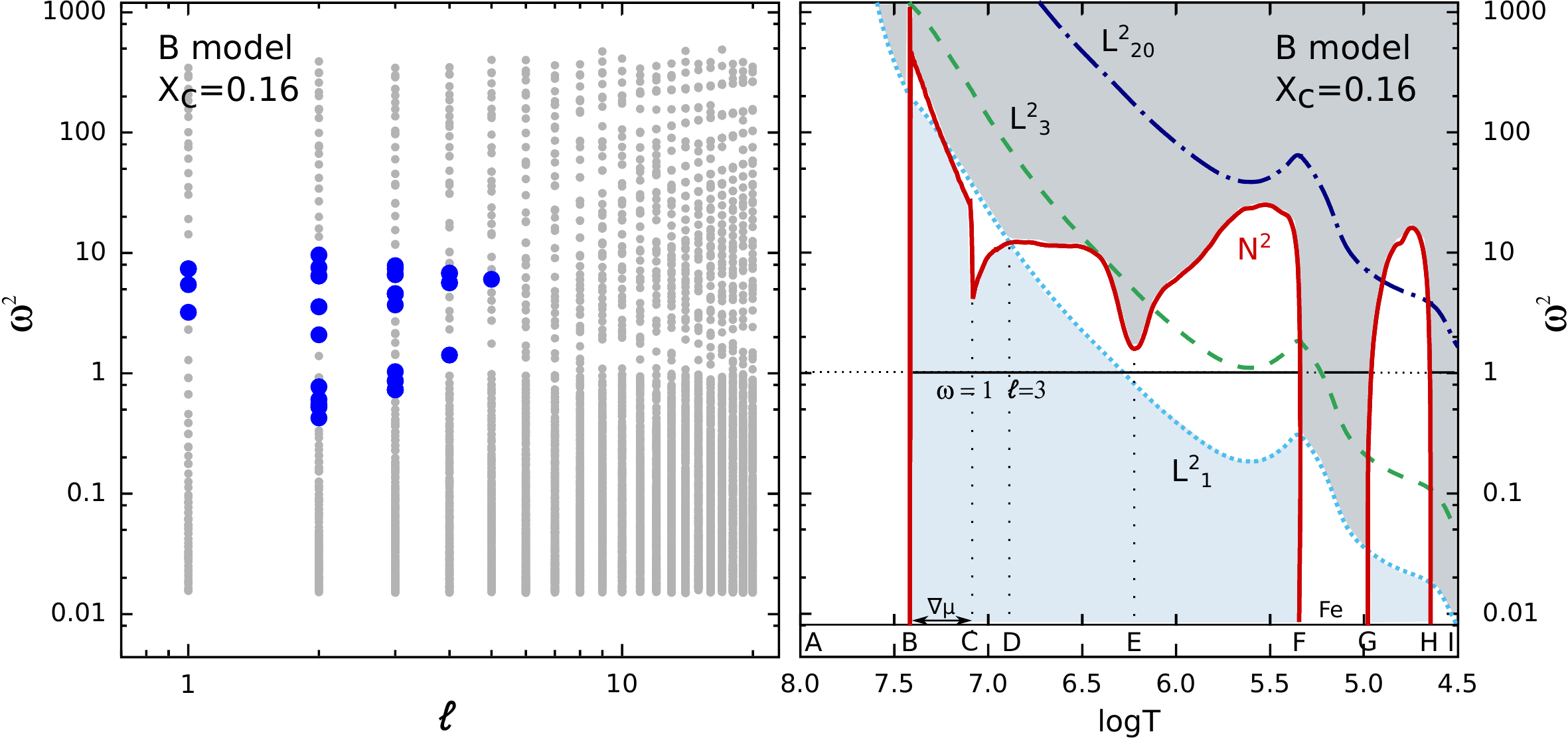}
\caption{
\textbf{Left} Modal diagram: range of dimensionless frequencies for a 50~${\rm M}_{\odot}$ MS model (labeled B in Tab.~\ref{tab_parameters}) as a function of the degree $\ell$ of the mode. Stable modes are represented by gray points while excited modes are shown with solid blue circles. \textbf{Right} Propagation diagram for the same model. 
The solid red line stands for the Brunt-V\"{a}is\"{a}l\"{a} frequency, and the Lamb frequency for three different degrees ($\ell$=1,3 and 20) is represented by the dotted, dashed and dot-dashed lines, respectively. The propagation zones for p and g modes with $\ell$=1 are highlighted by the dark gray and light blue shadow regions, respectively. The solid horizontal black line stands for an example of the propagation zones for a mode of $\omega=1.0$ with $\ell=3$. 
}
\label{fig_prop_diag_bv}
\end{figure*}

\section{Diagnostic diagrams}\label{sec_diagnostic}

In order to obtain a global picture of the stable and unstable modes for each computed model we used a version of the so-called modal diagrams\footnote{A modal diagram displays generally the frequency spectrum as a function of one global stellar parameter (e.g.  $\log T_{\rm eff}$, $\log g$)} in which the dimensionless frequency $\omega$ (or, alternatively, the frequency $f$) is displayed as a function of the degree $\ell$ of the mode. 

Left panel of Fig.~\ref{fig_prop_diag_bv} shows an illustrative modal diagram of this type for the particular case of a 50~${\rm M}_{\odot}$ MS model (see model B in Table~\ref{tab_parameters} for the global properties). Stable and unstable modes are represented by gray and big blue points, respectively. As discussed in the previous section, all the modal diagrams presented along this paper have been cleaned from the unphysical modes encountered in the computations (i.e., only the excited modes that have survived the visual inspection of the eigenfunctions are marked in blue).

The information provided by the modal diagrams is also complemented by analyzing the associated propagation diagrams. This second type of diagnostic diagrams allows us to identify the cavities in which p and g modes can propagate in a given model and for a certain value of $\ell$ \citep[e.g.][]{Unn89}. This is achieved by plotting together as a function of, e.g., $\log T$, the Lamb frequency
\begin{equation}
\label{eq_lamb}
L^{2}_{\ell}={\ell(\ell+1)\,c^{2}}/{r^{2}}
\end{equation}
(where $c$ is the sound speed and $r$ is the radius), and the Brunt-V\"{a}is\"{a}l\"{a} frequency 
\begin{equation}
\label{eq-bv}
 N^{2}=\frac{g_r}{r}\frac{d\ln P}{d\ln r} \, \bigg[\,\bigg(\frac{\partial \ln \rho}{\partial \ln T}\bigg)_{P,\mu}(\nabla_{ad}-\nabla)-\nabla_\mu\,\bigg(\frac{\partial \ln \rho}{\partial \ln \mu}\bigg)_{T,P}\bigg]
\end{equation}
(where $g_r$ is the local gravity, $P$, $\rho$, $T$, and $\mu$ are the pressure, density, temperature, and mean molecular weight, respectively, $\nabla_{ad}$ and $\nabla$ are the adiabatic and real temperature gradient, and $\nabla_\mu$ is the mean molecular weight gradient $\nabla_\mu=$ d$\ln\mu/$d$\ln P$).
This is illustrated in the right panel of Fig.~\ref{fig_prop_diag_bv}, which shows the propagation diagram corresponding to the model B mentioned above. The Lamb frequency (for  $\ell=1,3$ and $20$) and the Brunt-V\"{a}is\"{a}l\"{a} frequency are drawn with dashed and solid lines, respectively.
The propagation zones for p and g modes for the case $\ell$=1 are indicated by the dark gray and light blue areas, respectively, while the solid horizontal black line stands for an example of the propagation zones for a mode of $\omega$=1 with $\ell$=3. Generally speaking, those modes with $\omega^2>$ max$(N^2,L^2_\ell)$ are p modes and propagate in the more superficial layers, while those modes with $\omega^2<$ min$(N^2,L^2_\ell)$ correspond to g modes and propagate in the deeper layers. However, as illustrated by the horizontal line, there are also \emph{mixed} modes, which present both behavior depending of the regions of the stellar interior where they are propagating.
For example, the $\omega=1.0$, $\ell=3$ mode in Fig.~\ref{fig_prop_diag_bv} will have characteristics of p modes in the outer regions and of g modes in the inner one.

In summary, as described above, the combined information provided by the modal and the propagation diagrams shown in Fig.~\ref{fig_prop_diag_bv} allows to identify the frequency spectrum, the associated propagation cavity, and the mode type of excited modes in each considered stellar model.

\subsection{Some notes on the propagation diagrams}

For a better understanding of the following sections, we describe here in more details the propagation diagram for massive stars (Fig.~\ref{fig_prop_diag_bv}). We concentrate on the behavior of the Brunt-V\"{a}is\"{a}l\"{a} frequency since it plays a major role in the excitation of the instabilities computed in this paper. 
As shown in the propagation diagram, the Lamb frequency ($L_{\ell}$) is very large in the center and decreases monotonously towards the surface, following the decreasing c and the increasing r (see also Eq.~\ref{eq_lamb}). Its behavior is very similar from one model to another. 
On the contrary, the Brunt-V\"{a}is\"{a}l\"{a} frequency ($N$) can be very different depending on the specific mass and evolutionary state of the associated stellar model.
During the MS, $N^2 =0$ in the convective core (see Eq.~\ref{eq-bv}). In the layers surrounding the core, the presence of a $\nabla_\mu$ (due to the receding convective core during the evolution) produces a peak in the Brunt-V\"{a}is\"{a}l\"{a} frequency which intensity depends on the sharpness of the $\mu$ gradient \citep[e.g.][]{Noe10}. 
As a matter of fact, MS stars with intermediate masses (3 -- 20~${\rm M}_{\odot}$), usually present a unique peak in $N^2$(r) while the star evolves from the zero age main-sequence (ZAMS) due to the increasing $\nabla\mu$ region. 
Therefore, in that case, the star presents mainly 2 big cavities (p and g), and the frequency spectrum is characterized by avoided crossings between low-order p  and g modes during which the modes exchange their behavior.

In more massive stars, the behavior of $N^2$(r), and hence the propagation diagram, is more complex. In addition to the sharp peak due to the $\nabla\mu$ region on the MS,  the Brunt-V\"{a}is\"{a}l\"{a} frequency presents other bumps almost equally important leading to new cavities in which different types of modes may propagate. The propagation diagram (Fig.~\ref{fig_prop_diag_bv}, right panel) illustrates the most important characteristics of the Brunt-V\"{a}is\"{a}l\"{a} frequency in massive stars. Starting from the innermost region of the star (A), the actual limit of the convective core is shown by a first discontinuity in $N^{2}$ at $\log T$ $\sim$~7.4 (B). The subsequent decrease in $N^{2}$ (B-C) is due to the decrease of $g$ and $\mu$ (see Eq.~\ref{eq-bv} where $P/\rho\sim T/\mu$) in the region of variable $\mu$. It is followed by another discontinuity at $\log T \sim 7.1$ (C) once $\nabla\mu$ becomes 0.
This discontinuity corresponds to the maximal extension of the convective core during MS. 
Since above the $\nabla\mu$ region, $\nabla_{\rm rad}$ becomes smaller following the decreasing opacity, the shift between both temperature gradients in Eq.~\ref{eq-bv} gets larger and $N^{2}$ increases (C-D). This feature is discussed in more detail below. 
The next minimum in $N^{2}$ is produced by the deep opacity bump at $\log T \sim 6.25$ due to the L-transition of iron (E). One can also see the convective zone due to the iron opacity bump at $\log T \sim 5.2$ (F-G) and a superficial convective zone due to the ionization of helium ($\log T \sim 4.6$, H-I). 

Apart from the large deep opacity bump, the most important characteristic of the $N^{2}$ frequency in massive stars is related to the sharp feature produced by the important radiation pressure around $\log T \sim 7.0$ (C-D). 
Actually, the increasing contribution of the radiative pressure to the total pressure (with mass, and during the evolution on the MS, due to the increasing central temperature) induces a decreasing adiabatic temperature gradient. Since, both gradients ($\nabla_{\rm rad}$ and $\nabla_{\rm ad}$) are already close to one another, the $\nabla_{\rm rad}$ becomes more easily larger than $\nabla_{\rm ad}$, leading eventually to the appearance of an intermediate convective zone (ICZ) on the post-MS (see Sect.~\ref{sub_bi_l_equal1}). 
However, even before the onset of an ICZ, the discontinuity in the Brunt-V\"{a}is\"{a}l\"{a} frequency (C), already present during MS at the upper limit of the $\nabla_{\mu}$ region, shows a minimum value smaller and smaller as the mass increases due to the increasing radiation pressure leading to a smaller coefficient $(\partial \ln \rho / \partial \ln T)_{P,\mu}$ (Eq.~\ref{eq-bv}). This lower minimum value in the Brunt-V\"{a}is\"{a}l\"{a} frequency will have a similar effect as the ICZ, as we will see in Sect.~\ref{sub_bi_l_larger1}.

\section{Instability strips}\label{sec_instabilitystrips}

\begin{figure}[!t]
\centering
\includegraphics[width=0.49\textwidth]{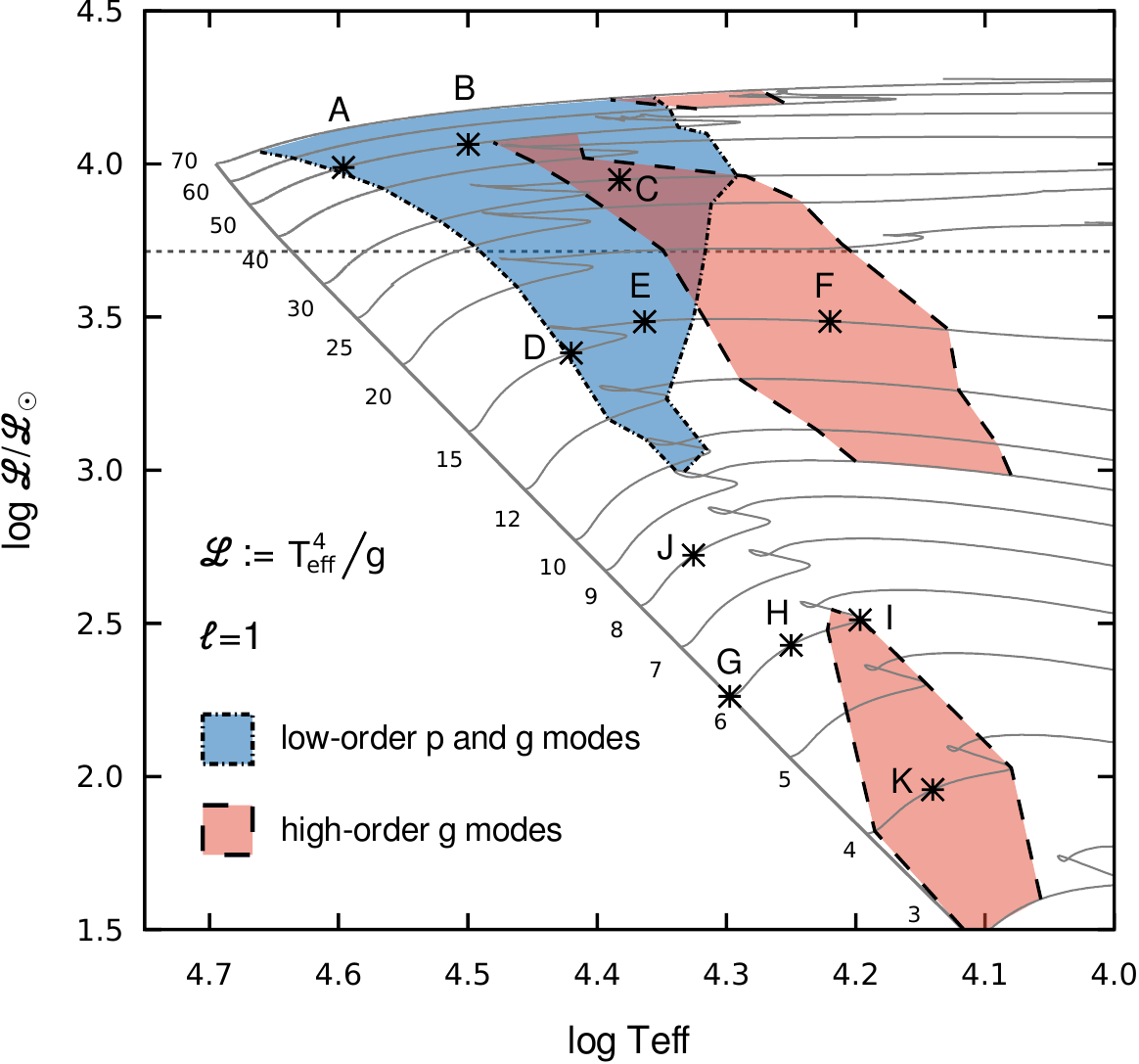}
\caption{Instability domain predictions in the sHR diagram for dipole modes ($\ell=$~1, see Sect.~\ref{sub_bi_l_equal1}). P-mode instabilities are represented by the blue area and the g-mode domain is shown in red.  
The ATON, non-rotating evolutionary tracks are over plotted. The horizontal dashed line indicates the limit above which strange mode oscillations are most likely expected. Black asterisks stand for 9 selected models labeled A to I which will be more deeply investigated in Sect.~\ref{sub_bi_l_larger1}. The global properties of these models are given in Tab.~\ref{tab_parameters}.
}
\label{fig_inst_l1}
\end{figure}

\begin{figure}[!t]
\centering
\includegraphics[width=0.49\textwidth]{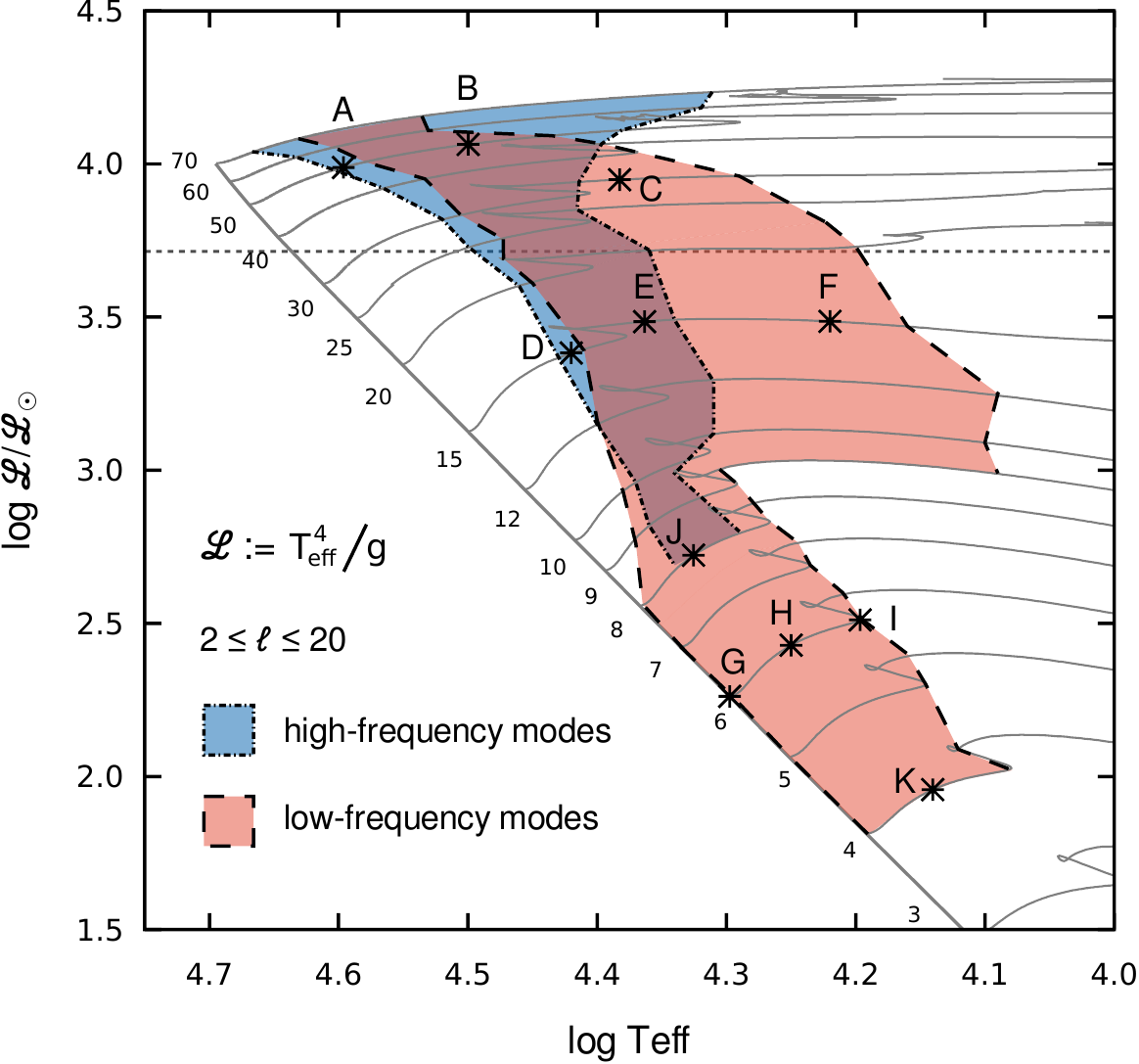}
\caption{Same caption as in Fig.~\ref{fig_inst_l1} except that the instability domain predictions are now done for 2~$\le\ell\le$~20 pulsation modes (see Sect.~\ref{sub_bi_l_larger1}).
}
\label{fig_inst_lgtab_parameters}
\end{figure}

\begin{table} [tb]
\caption[]{Global stellar parameters of 11 selected models: mass, effective temperature, gravity, the quantity $\mathscr{L}$ defined as $T_{\rm eff}^4/g$, central hydrogen abundance, and dynamical time in hour. See the location of these models in the spectroscopic HR diagram in Figs.~\ref{fig_inst_l1} and \ref{fig_inst_lgtab_parameters}.
}
\label{tab_parameters}
\centering
\begin{tabular}{cccccccccc}
\hline\hline
& $M/{\rm M}_{\odot}$  &  $\log T_{\rm eff}$  & $\log g$ & $\log \mathscr{L}/\mathscr{L}_{\odot}$ & $X_c$ & $\tau_{\rm dyn}$ 	\\
\hline
A  	&       50             &    4.60          &     3.79                  &     3.99   	   &  	0.363	&	20.8		\\
B   	&       50             &    4.50          &     3.33                  &     4.06   	   & 	0.161 	&	44.3		\\
C  	&       30             &    4.38          &     2.97                  &     3.95   	   &  	0.000	&	76.3		\\
D  	&       15             &    4.42          &     3.69                  &     3.38   	   &  	0.140	&	19.9		\\
E  	&       15             &    4.36          &     3.36                  &     3.49   	   & 	0.000 	&	34.8		\\
F  	&       15             &    4.22          &     2.79                  &     3.49   	   &  	0.000	&	93.9		\\
G  	&       6             &     4.30         &      4.32                 &      2.26  	   &  	0.697	&	5.34		\\
H  	&       6             &     4.25         &      3.96                 &      2.43  	   &  	0.303	&	9.88		\\
I  	&       6             &     4.20         &      3.67                 &      2.51  	   &  	0.035	&	16.5		\\
J  	&       8             &     4.33         &      3.97                 &      2.72  	   &  	0.331	&	10.5		\\
K  	&       4             &     4.14         &      4.00                 &      1.96  	   & 	0.316 	&	8.45		\\
\hline
\end{tabular}
\end{table}

\begin{figure*}[!t]
\centering
\includegraphics[width=0.8\textwidth]{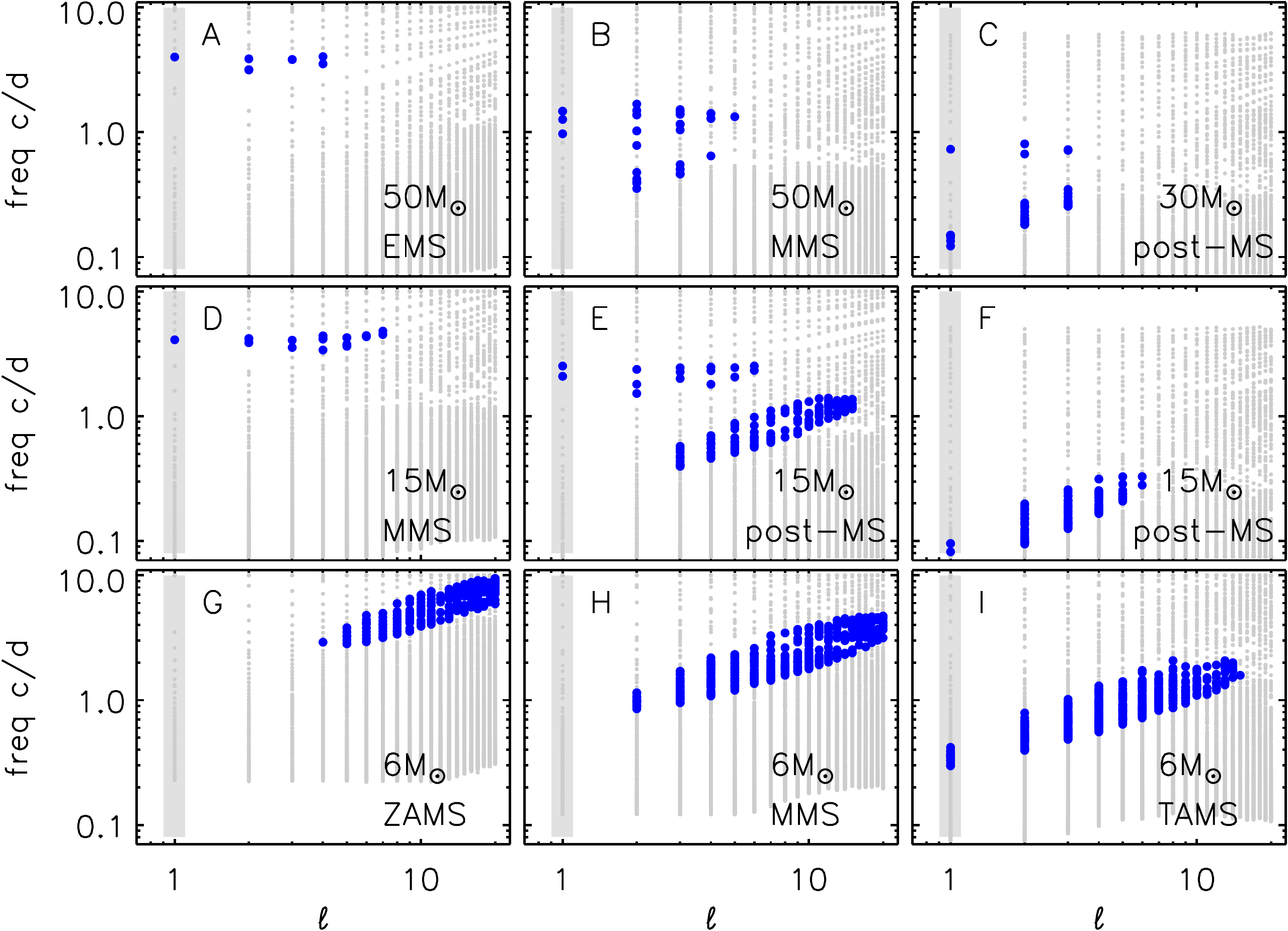} 
\caption{Range of frequencies in cycle per day (c/d) for models A to I. Gray dots stand for the frequencies of the stable modes computed  whereas excited mode frequencies are represented by blue bigger dots. From left to right, the rows represent the evolution from the MS to the post-MS: models A to C for high masses (30 -- 50~${\rm M}_{\odot}$),  models D to F for 15~${\rm M}_{\odot}$ and models G to I for 6~${\rm M}_{\odot}$ MS models only (see Fig.~\ref{fig_inst_l1} and Tab.~\ref{tab_parameters}). Rough evolutionary states are indicated for each model in the right bottom corner (ZAMS, EMS, MMS, TAMS and post-MS stand for zero age MS, early MS, middle of MS, terminal age MS and hydrogen burning shell models). 
Note that the detectability of the pulsation modes in photometry is limited to the low degree ($\ell$ up to 3 -- 4) whereas higher degree modes should be observed thanks to spectroscopy.}
\label{fig_models_AI}
\end{figure*}

\begin{figure}[!t]
\centering
\includegraphics[width=0.3\textwidth]{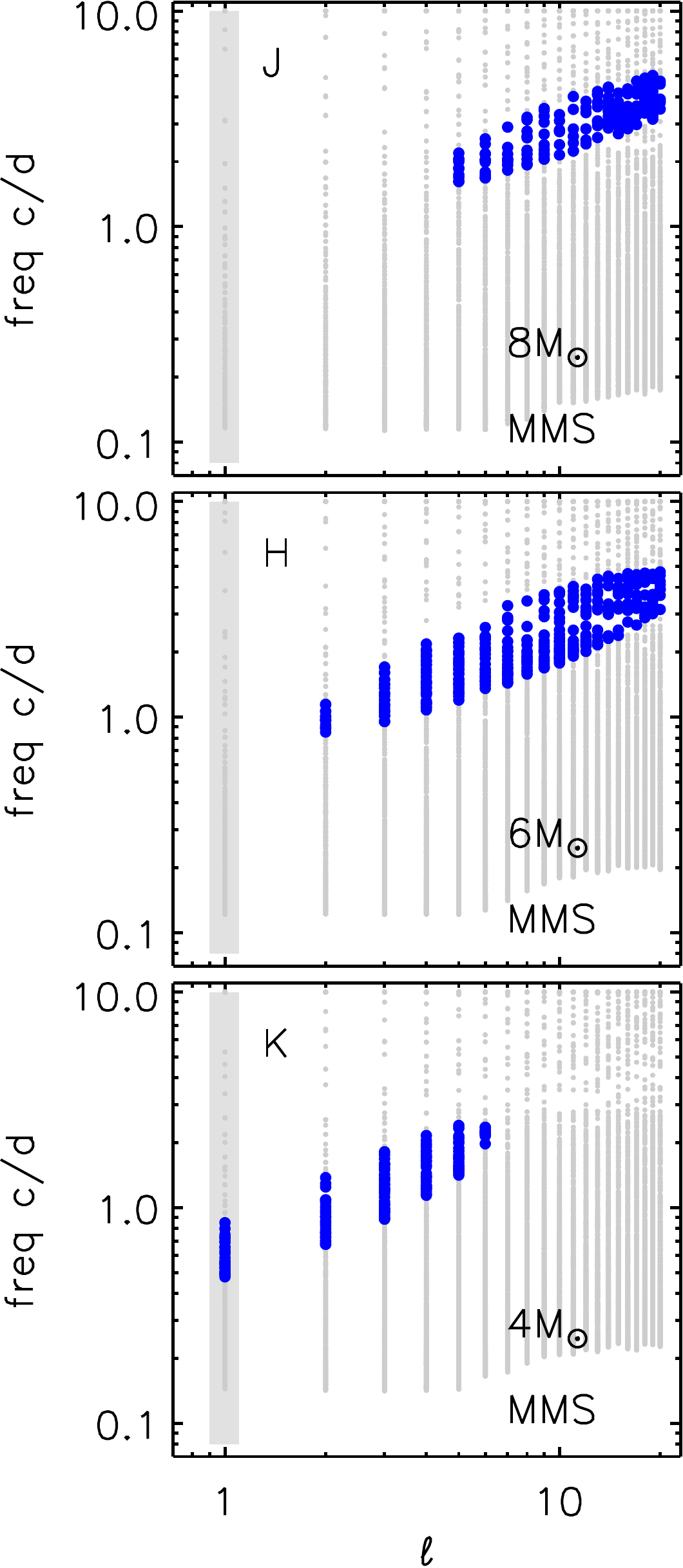}
\caption{Same caption as in Fig.~\ref{fig_models_AI} but for models J to K, selected, from top to bottom, at the same evolutionary state ($X_c\sim 0.3$) but with decreasing masses from 8 to 4~${\rm M}_{\odot}$ (see Fig.~\ref{fig_inst_l1} and Tab.~\ref{tab_parameters}).}
\label{fig_models_JK}
\end{figure}

By investigating modal diagrams, we have determined the instability boundaries of pulsations due to the $\kappa$-mechanism in the iron opacity bump. These results, along with the ATON, non-rotating evolutionary tracks are presented in the spectroscopic Hertzsprung-Russell diagram \citep[sHR diagram, ][]{Lan14, Cas14}, separating the case of dipole modes ($\ell=1$, Fig.~\ref{fig_inst_l1}) and higher-degree modes ($2\le \ell \le 20$, Fig.~\ref{fig_inst_lgtab_parameters}). These figures are complemented with some illustrative cleaned modal diagrams for 11 selected models, labeled from A to K (Figs.~\ref{fig_models_AI} and \ref{fig_models_JK}). The location of these models in the sHR diagram are shown by black asterisks in Figs.~\ref{fig_inst_l1} and \ref{fig_inst_lgtab_parameters} and their main parameters are summarized in Table~\ref{tab_parameters}. In this case, we have used the dimensional frequency $f$ in cycles per day (instead of the dimensionless frequency $\omega$ as in Fig.~\ref{fig_prop_diag_bv}) for a better utility of these diagrams from an observational point of view. The values of $\tau_{\rm dyn}$ (in hours) provided in Table~\ref{tab_parameters} for each model can be used to transform $f$ into the dimensionless variable $\omega$.

The panels of Figs.~\ref{fig_models_AI} and~\ref{fig_models_JK} are organized following the location of the selected models in the sHR diagram. In Fig.~\ref{fig_models_AI}, every row displays models with similar masses (except for model C; decreasing masses from the top to the bottom row), and, the models are evolving from the beginning of the MS to the post-MS from left to right (i.e. every column corresponds to a given evolutionary state). Rough evolutionary states are indicated for each model in the right bottom corner of the modal diagram (ZAMS, EMS, MMS, TAMS and post-MS stand for zero age MS, early MS, middle of MS, terminal age MS and hydrogen burning shell models). Finally, Fig.~\ref{fig_models_JK} represents models with different masses (8 to 4~${\rm M}_{\odot}$ from the top to the bottom panel) selected at the same evolutionary stage (MMS, $X_c=0.3$).

Modes appear into 2 groups depending on their frequencies (in particular, see models B, C and E in Fig.~\ref{fig_models_AI}): 
\begin{itemize}
 \item modes with high frequencies (small periods) behave mainly like p modes, these modes are excited with low orders;
 \item modes with low frequencies (large periods) have a general g-mode behavior, and these modes are usually excited with high orders.
\end{itemize}
The limit between these two kinds of modes depends on the frequency but also on the degree $\ell$ of the mode. For this reason, we do not use the usually accepted criterion in dimensionless frequencies of $\omega>1$ ($\omega<1$) for p modes (resp. g modes), and we mainly refer in the following to the high-frequency and low-frequency modes, especially when discussing the results for $\ell>1$. 

For a better understanding of the instability domain predictions, we first concentrate on the results for $\ell=1$, and then extend the description of the results to higher-degree modes (2 $\le\ell\le$ 20), for which the physics discussed in the $\ell=1$ case applies too.

\subsection{Instability domains for dipole modes ($\ell=1$)}\label{sub_bi_l_equal1}

The results for dipole modes ($\ell=1$) are presented in the sHR diagram in Fig.~\ref{fig_inst_l1} while the associated modal diagrams for the selected models A--K are highlighted with a vertical gray band in Figs.~\ref{fig_models_AI} and \ref{fig_models_JK}. We separate the discussion below into the three major instability domains.  

We start by discussing the lower mass regime (from 3 to 6~${\rm M}_{\odot}$) for which instabilities are found for MS stars only. These models are characterized by high-order g modes only (represented by the red area limited by dashed lines in Fig.~\ref{fig_inst_l1}), which basically maps the SPB instability strip for $\ell$=1. The periods related to these modes are of $1 - 3$ days (see panels I and K in Figs.~\ref{fig_models_AI} and \ref{fig_models_JK}, respectively). 

Moving up in the sHR diagram, we find low-order p and g modes, which are predicted on the MS and also on the early post-MS of massive models. These modes are represented by a blue area limited by dotted lines in Fig.~\ref{fig_inst_l1}.
In the intermediate mass range ($\sim$ 10 -- 30~${\rm M}_{\odot}$), these excited modes with periods of the order of $6$ hours constitute the $\beta$ Cephei instability strip (e.g., model D, $\ell=1$). 
For the models with larger masses (M~$\ge$ 30~${\rm M}_{\odot}$),  some adiabatic strange modes are also excited with periods of a few hours (models A and B), in addition to the low order p and g modes. These adiabatic strange modes are excited in the iron opacity bump due to the $\kappa$-mechanism and their strange mode behaviour is related to the amplitude enhancement of the mode trapped into a superficial cavity \citep[e.g.][]{Saio1998,Gla09}. The nature of these modes has been confirmed by investigating the eigenfunctions.
These instabilities extend to the post-MS phase, especially for models with masses up to 40~${\rm M}_{\odot}$ (with dimensionless frequency $\omega\sim 3$, i.e. periods of the order of hours). 
In particular, in these post-MS models, the instabilities are essentially due to low-order g modes that present a mixed mode behavior, i.e. they propagate into the central layers as g modes and they have a p mode behavior in the superficial layers. 

Finally, a large number of high-order g modes are expected to be excited mainly during the post-MS phase of stars with masses larger than 9~${\rm M}_{\odot}$ (red area in Fig.~\ref{fig_inst_l1} and models C and F in Fig.~\ref{fig_models_AI}). These modes have periods of about $\sim 5 - 15$ days, and this region of the sHR diagram is often referred to as the extension of the SPB instability strip to higher masses. 
They are predicted in models which present an ICZ surrounding the radiative core. The ICZ acts as a barrier and prevents the modes from entering the damping core.
Indeed, the density contrast between the helium core and the envelope is very high in massive post-MS models. As a result, the Brunt-V\"{a}is\"{a}l\"{a} frequency reaches huge values in the core and a large radiative damping occurs \citep[the radiative damping is proportional to $N^2$, see Eq.~3 in][]{Dupret2009}.
However, the ICZ, which develops above the radiative core during post-MS evolution, reflects the modes and allows therefore a sufficient excitation by the $\kappa$-mechanism in the iron opacity bump \citep{Saio2006,Gau09,Godart2009}, similarly to the excited modes in SPB stars.  
In our models, the ICZ appears in the models of 12~${\rm M}_{\odot}$ (up to higher masses) although g-mode instabilities are already predicted in models of 9~${\rm M}_{\odot}$. Actually, we found here that the sharp feature in the Brunt-V\"{a}is\"{a}l\"{a} frequency discussed in Sect.~\ref{sec_diagnostic} (at $\log T \sim 7.1$ in Fig.~\ref{fig_prop_diag_bv}) already produces the reflection needed for exciting the g modes if the minimum is low enough, in agreement with \cite{Das13}, as it is the case for our models with M~$\ge$ 9~${\rm M}_{\odot}$. 

The instability strip for these high-order g modes (low frequencies) overlaps with the low-order p and g modes instability strip (high frequencies) for the most massive models.  The overlapping of both areas appears in purple in Fig.~\ref{fig_inst_l1}. Mainly located on the post-MS (for $\sim$ 15 -- 30~${\rm M}_{\odot}$), it reaches the MS for masses of 40 -- 50~${\rm M}_{\odot}$.
When both these frequency ranges are excited at the same time (as it is the case at the end of the MS for a 40~${\rm M}_{\odot}$ star for example, for which the frequency spectrum is similar to model C), a stable region lies between the frequency domains of excited modes (Fig.~\ref{fig_models_AI}). The amplitude of the modes in that range of intermediate radial orders are larger in the core than in the envelope, and the important radiative damping prevents the $\kappa$-mechanism to be efficient enough. 
Two effects explain this behavior: (1) the evanescent zone is larger in this frequency range which leads to a larger coupling between p and g cavities and thus to a larger radiative damping \citep{Dupret2008} and (2) the amplitudes of the eigenfunctions in the superficial layers
show a minimum in this frequency range \citep{Dziembowski1993}.

All these theoretical predictions for the dipole instability domains are in good agreement with the work of \citet{Saio2011}. 
For a more extensive description of the properties of the various theoretical instability domains for massive stars for low degree $\ell$ and for modes excited by other mechanism than the opacity mechanism we refer the reader to \citet[][]{Saio2011,Saio2015a,Saio2015} and the reviews of \citet[][]{Godart2014,Samadi2015}.

\subsection{Instability domains for $~2~\le\ell\le~20$}\label{sub_bi_l_larger1}

As stated in Sect.~\ref{sec_intro}, the motivations that guided us through this work are the investigation of the physical origin of the macroturbulent broadening and the interpretation of the observed line-profile variability of OB stars. In this context, we extended our computations to high degree modes ($\ell >$ 3), which signatures can be visible in spectroscopy, though they are not detectable in photometry. 
The instability domains predicted for $\ell$=~2 to 20 are shown in Fig.~\ref{fig_inst_lgtab_parameters} by shaded areas.  They are complemented by the information contained in the modal diagrams (as illustrated in Fig.~\ref{fig_models_AI} and \ref{fig_models_JK}). 

We divided our predictions for the frequency ranges and instabilities into two groups. First, a  group of modes with high frequencies (small periods), 
for which $\omega \sim 2-3$, is represented by the blue area limited by dotted lines in Fig.~\ref{fig_inst_lgtab_parameters}. This group of pulsations presents mostly a p-mode behavior and constitutes a rather sparse and discrete spectrum (see modes with f $\sim$ 4.0 c/d for models A and D, $f \sim$ 1.5-2 c/d for models B and E, and modes with f $\sim$ 0.8 c/d for model C).  The periods associated to these modes correspond to $\sim 8$ hours for models A and D, $\sim 12$ h for model B and E and $\sim 24$ h for model C, with increasing periods along the evolution. As for dipole modes, these modes are excited with low orders. 
Secondly, a group of lower frequencies (large periods), for which $\omega\sim$1, is represented in the red area with dashed contour lines. These pulsations mainly present a g-mode behavior and are mainly excited with high orders. These modes constitute a much denser spectrum (see the lower frequency group in models B, C, E, F, G, H, I). The periods range from a couple of hours to $10$ days.
Usually well detached, these two groups of modes reach a roughly common frequency range in the most massive models (see model B), in which the excited frequencies from p and g modes are found very close to each other and the modes behave as mixed modes, in agreement with \citet{Bal99} (see the explanation in Sect.~\ref{sub_bi_l_equal1}).  

As expected, the instability domains cover a wider region of the sHR diagram when considering high-degree pulsations.
This effect is noticed in low-order p and g mode instabilities, for which the instability domain mainly widens towards smaller masses and reaches MS models of 8~${\rm M}_{\odot}$ (rather than 10~${\rm M}_{\odot}$ for dipole modes). It is also marked when considering the high-order g-mode domain. Indeed, the instability bands reach hotter $T_{\rm eff}$ in all the instability region. In particular, the post-MS gap between low-order p and g modes and high-order g modes is now completely filled in by excited modes in models from 9 to 12-15~${\rm M}_{\odot}$. 
In these models, as for example in model E, g modes are only predicted when considering $\ell \ge 3$ which would, in this case, make them difficult to be observed in photometry.
Furthermore, theoretical computations predict now excited g modes in the high-mass models (up to 70~${\rm M}_{\odot}$). See for example model B, for which excited quadrupole modes should be detectable in photometry (P $\sim 2.5$ days), though g modes are not predicted for $\ell=1$.  Finally, intermediate mass models (from 5 -- 12~${\rm M}_{\odot}$) now present instabilities on the MS for 5 to 10~${\rm M}_{\odot}$. These modes have quite large frequencies -- in regards of their g-mode behavior -- (small periods reaching the few hours) due to their mixed mode behavior. Here again, while models G and J are characterized by the excitation of a dense spectrum of g modes, these are hardly expected to be detected with photometry while the situation is more favorable for models H, I and K.
These results show that with the increasing quality of the photometric datasets offered by the space missions, it should become more and more possible to detect intermediate-degree g modes in some MS massive stars.

Some regions of the sHR diagram remain however completely stable when considering $\ell=$ 1 to 20 modes: the post-MS of models with 3 -- 8~${\rm M}_{\odot}$, the region close to the ZAMS of 9 -- 70~${\rm M}_{\odot}$ models, and the much evolved models of 9 -- 70~${\rm M}_{\odot}$.
In particular, for this latter group, it is important to remark that numerous modes are found positive to an instability for models with $M > 20\,{\rm M}_{\odot}$ and with  $\log T_{\rm eff} < 4.3$. The eigenfunctions of these modes are however sometimes ambiguous and more investigation needs to be performed in order to classify the types of modes encountered (including strange modes\footnote{For reference, the commonly used lower boundary for the appearance of strange modes ($L/M\sim$10$^4$, in solar units) is indicated with an horizontal line in Fig.~\ref{fig_inst_lgtab_parameters}.}) and to identify the excitation mechanisms. Therefore, although we have computed the instabilities until $\log T_{\rm eff}=4.0$ and found excited modes in that region, we have decided to restrain the instability domains to the modes excited by the $\kappa$-mechanism due to the iron opacity bump only. Note that the spectra of the excited modes encountered in these models were usually less dense than for the earlier evolution (various modes were found in all the frequency range, but less concentrated in a given domain of frequency). 

Last, we remark that the frequency spectrum for lower masses reaches lower-degree modes for decreasing masses at the same evolutionary state (see models J, H to K), and that no modes with $\ell>$1 are predicted for 3~${\rm M}_{\odot}$. We discuss the global behavior of the high-degree modes when the stars evolve from the MS to the post-MS in the next section. 

\begin{figure}[!t]
\centering
\includegraphics[width=0.48\textwidth]{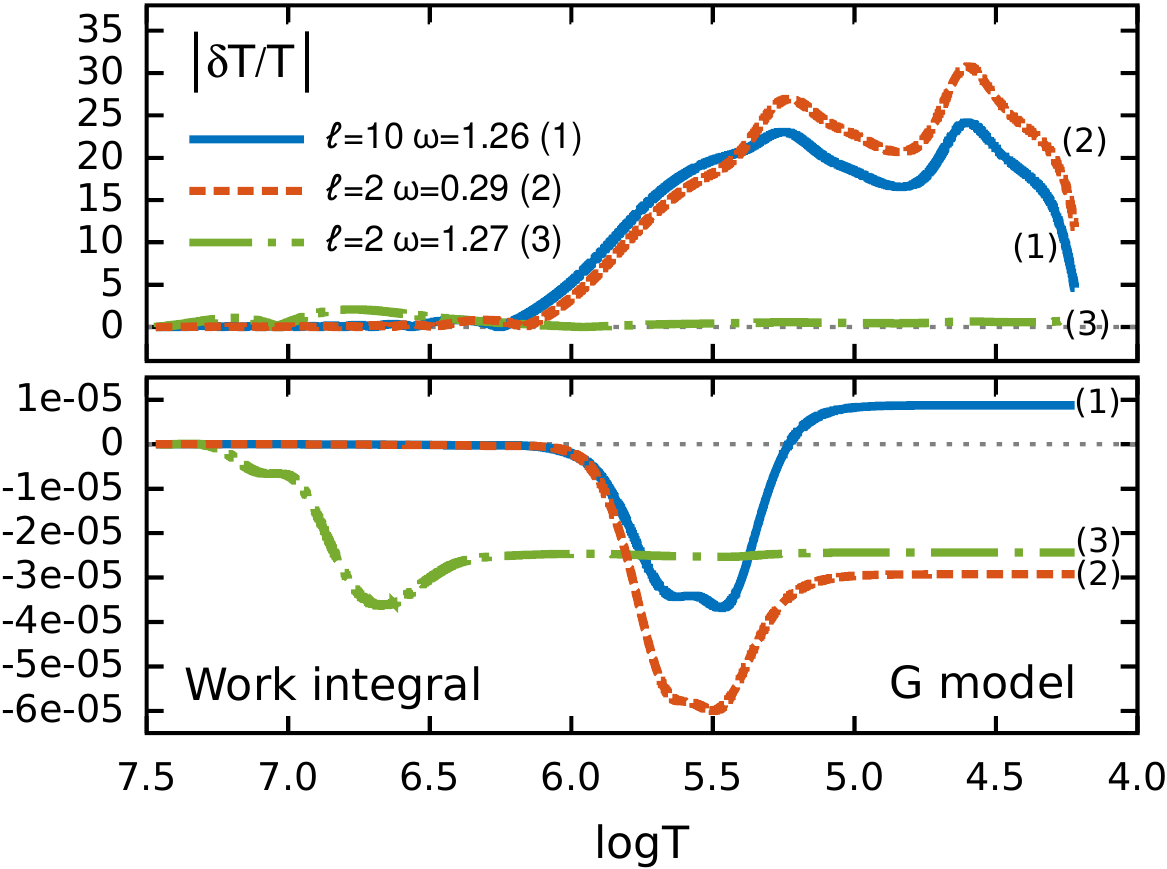} 
\caption{Perturbed temperature (top panel) and work integral (bottom panel) for three modes selected in model G: one excited mode (see the positive work integral at the surface on the bottom panel) with $\omega=1.26$ and $\ell=10$ (blue solid line) and two stable quadrupole modes with $\omega=1.27$ (dot-dashed green line) and $\omega=0.29$ (dotted orange line).}
\label{fig_ef}
\end{figure}

\subsection{Properties of high-degree modes}\label{sub_global_l}

A global look at modal diagrams presented in Figs.~\ref{fig_models_AI} and \ref{fig_models_JK} allows us to characterize the behavior of the high-degree pulsations with the stellar evolution for different initial mass. For $M\,\gtrsim 9\,{\rm M}_{\odot}$, no pulsation are found close to the ZAMS. Modes with $\ell=1$\,--\,10 and high frequencies (i.e., having a p-mode behavior) are the first to appear at some point during the MS -- the earlier the more massive the star --. Then, as the stars evolve towards lower $T_{\rm eff}$ in the sHR diagram (i.e. looking from model A to C for higher masses and from model D to F for 15~${\rm M}_{\odot}$), a large group of modes with lower frequencies (g-mode behavior) appears with a large range of degree $\ell$, and the number of excited p modes diminishes (often for all degrees at the same time). As a general trend, the degree $\ell$ of g modes decreases along the evolution, even reaching $\ell=1$ in the post-MS region. Last, the red part of the sHR diagram is characterized by other types of modes which are not taken into account here as already mentioned in Sect.~\ref{sub_bi_l_larger1}.

Concerning lower masses (models G to I), the ZAMS is characterized by SPB type modes with high degrees (modes with $\ell$ up to 20 are excited, see model G and H). The degree of the mode, along with the frequency range of the excited modes are then decreasing over the MS evolution. 
We note that stars with masses between $\sim$7\,--\,9~${\rm M}_{\odot}$ predicted to pulsate with g modes only present modes of degrees $\ell$ larger than 3.
Finally, models J, H and K, selected at a given state of evolution ($X_c\sim$ 0.3, Fig.~\ref{fig_models_JK}),  all present SPB type g modes.  At the same evolutionary stage, the largest mass models present high-degree excited modes with periods of the order of hours (model J and H present excited modes with $\ell$ up to 20). By decreasing the mass of the models (from J to K), we found modes excited with a decreasing $\ell$, and the frequency range of instability reaches the usual long periods of the SPB stars observed in photometry, of the order of days (model I). The post-MS phase of evolution is then characterized by pure stable models. 

While high-frequency modes appear and disappear similarly altogether during the evolution, we note 2 general trends concerning the modes with the lowest frequencies (`g modes') for increasing $\ell$:
\begin{itemize}
 \item the range of unstable g modes is moving towards higher frequencies;
 \item the blue boundary of the instability domain of g modes is shifted towards warmer $T_{\rm eff}$, overlapping the instability of $\beta$ Cephei stars.
\end{itemize}
These 2 trends can be explained with the following reasoning, based on the 2 conditions needed for an efficient $\kappa$-mechanism.
On the one hand, in order to have an efficient $\kappa$-mechanism, the amplitudes of the g modes have to be large enough in the opacity bump compared to the g cavity where the radiative damping plays an important role. Therefore, since the amplitudes of the eigenfunctions for g modes are essentially a function of the ratio $\omega^2/(\ell(\ell+1))$ (in the asymptotic limit, a fixed $\omega^2/(\ell(\ell+1))$ corresponds to a given radial order), the condition on the amplitudes implies therefore a given range for the $\omega^2/(\ell(\ell+1))$ ratio. 
On the other hand, an efficient $\kappa$-mechanism also requires that the transition zone (the region in the star where the pulsation period is of the same order as the thermal relaxation time) 
coincides with the region of the opacity bump. This is the case for a given range of frequencies, independently of the degree $\ell$ of the mode. 
These two conditions generate therefore 2 constraints, one depending on the degree of the mode and the other not. The intersection between these constraints defines the group of unstable g modes in the modal diagrams. 
To illustrate this effect, we selected 3 modes in model G:
\begin{itemize}
 \item (1) a mode with a large degree ($\ell=10$) with $f=5.67$ c/d (dimensionless frequency $\omega = 1.26$) which has been selected among the group of unstable modes,
 \item (2) a stable mode with a low degree ($\ell=2$), and with a lower frequency ($f$=1.29 c/d and $\omega = 0.29$), which has been selected to fit the same $\omega^2/(\ell(\ell+1))$ ratio as mode (1), and
 \item (3) a stable quadrupole mode with the same frequency as mode (1) ($f$=5.70 c/d and $\omega = 1.27$).
\end{itemize}
Fig.~\ref{fig_ef} shows the eigenfunction of the perturbed temperature (top panel) and the work integral (bottom panel) for these 3 modes (the blue solid line stands for mode (1), the dotted orange line for the stable mode (2) and mode (3) is represented by the dot-dashed green line).
From Fig.~\ref{fig_ef}, we find that mode (1) presents large amplitude in the superficial layers and a positive work at the surface: it is excited. However, for a lower-degree mode with the same frequency (as it is the case for mode (3)), the eigenfunction takes low values in the superficial layers due to the larger evanescent zone \citep{Dupret2008}, and because of the
boundary condition: $\delta P/P = ( l(l+1)/\omega^2 - 4 - \omega^2) \,\,\delta r/R$ \citep{Buta1979}. As a result, the mode stays stable. 
Moreover, for a given radial order (i.e. for a fixed $\omega^2/(\ell(\ell+1))$, see modes (1) and (2)), the eigenfunctions of the perturbed temperature present large amplitudes in the superficial layers, and, especially, in the region of the opacity bump due to the iron elements, favoring an efficient $\kappa$-mechanism. However, unlike mode (1), mode (2) is found stable. Indeed, its frequency being lower, the transition region corresponding to mode (2) is situated deeper into the star. As a result, it is located outside the opacity bump of iron (characterized by a fixed $\log T$), preventing the excitation of the associated mode. 
In short, the layout of the unstable low-frequency modes in the modal diagrams depends on 2 conditions controlling the  efficiency of the $\kappa$-mechanism, i.e. the constraint on the amplitudes of the modes depending on $\ell$, and the constraint on the location of the transition region, which depends on the frequency itself only: a larger $\ell$ corresponds therefore to bluer instability domains and to excited modes of higher frequencies. 

Before confronting our theoretical predictions to spectroscopic observations in the next section, we remark that the computations performed here are based on a given set of input physics (described in Sect.~\ref{sec_computations}) which affects the exact boundary of the instability domains in the sHR diagram. We will refer to this in more details in the context of macroturbulence in Sect.~\ref{sec_macro}.

\section{Link with observations: macroturbulent broadening}\label{sec_macro}

The most commonly considered scenario to explain the existence of a non-rotational line-broadening mechanism -- traditionally quoted as macroturbulent broadening -- shaping the line profiles of O stars and B~Supergiants claims that this broadening is a  signature of the collective effect of stellar oscillations \citep{Luc76, Aer09}.
While there are some empirical hints supporting this hypothesis \citep{Sim10, Sim16, Aer14}, the exact physical mechanism driving the type of instabilities required to produce the observed line profiles is still under debate. One obvious possibility refers to heat-driven modes produced by a $\kappa$-mechanism. However, other options such as stochastically excited oscillations driven by turbulent pressure fluctuations initiated either in sub-surface convective zone \citep{Gra15} or by the interface between the convective core and the radiative envelope \citep{Rog13,Aer15} have been also recently proposed as potential hypotheses. Actually, as commented by \cite{Sim16}, different excitation mechanisms could be  contributing, at the same time and to some degree, to the global broadening of the line profiles depending on the mass and the evolutionary stage of the star.

In this section we connect the observations gathered in the framework of the IACOB project \citep{Sim15b} to investigate the physical origin of the so-called macroturbulent broadening in the O and B star domain with our predictions for instability domains of high-degree modes driven by the $\kappa$-mechanism. This is  a continuation of the work started in \cite{Sim16}, where we compared the empirical information extracted from a sample of high-resolution single-snapshot spectra of $\approx$~430 Galactic O and B stars to the predictions for the dipole instability domains presented in this paper. 
We found an important number of stars in the upper part of the sHR diagram having a dominant contribution of the macroturbulent broadening, but located outside the g-mode instability domains. Furthermore, the O stars and B~Sgs located inside the instability domains populate different regions in the diagram characterized by different type of excited modes. However, a quite homogeneous distribution of profile types (in terms of global shape and relative contribution between the macroturbulent and the rotational broadening) is found for all these stars, despite the expected different effect on the line profiles due to the specific combination of excited modes. All together, these results could be interpreted as a strong empirical challenge to macroturbulent broadening in O stars and B~Sgs being a spectroscopic signature of only non-radial modes excited by the $\kappa$-mechanism. However, as commented in \cite{Sim16}, the situation may improve when including the predictions for high-degree modes or when modifying the input parameters in the stellar model computations, such as the opacities and the metal mixture. These two possibilities are discussed in detail in the next sections.

\subsection{High-degree modes vs. macroturbulent broadening}\label{sec_macro_shr}

\begin{figure}[!t]
\centering
\includegraphics[width=0.49\textwidth]{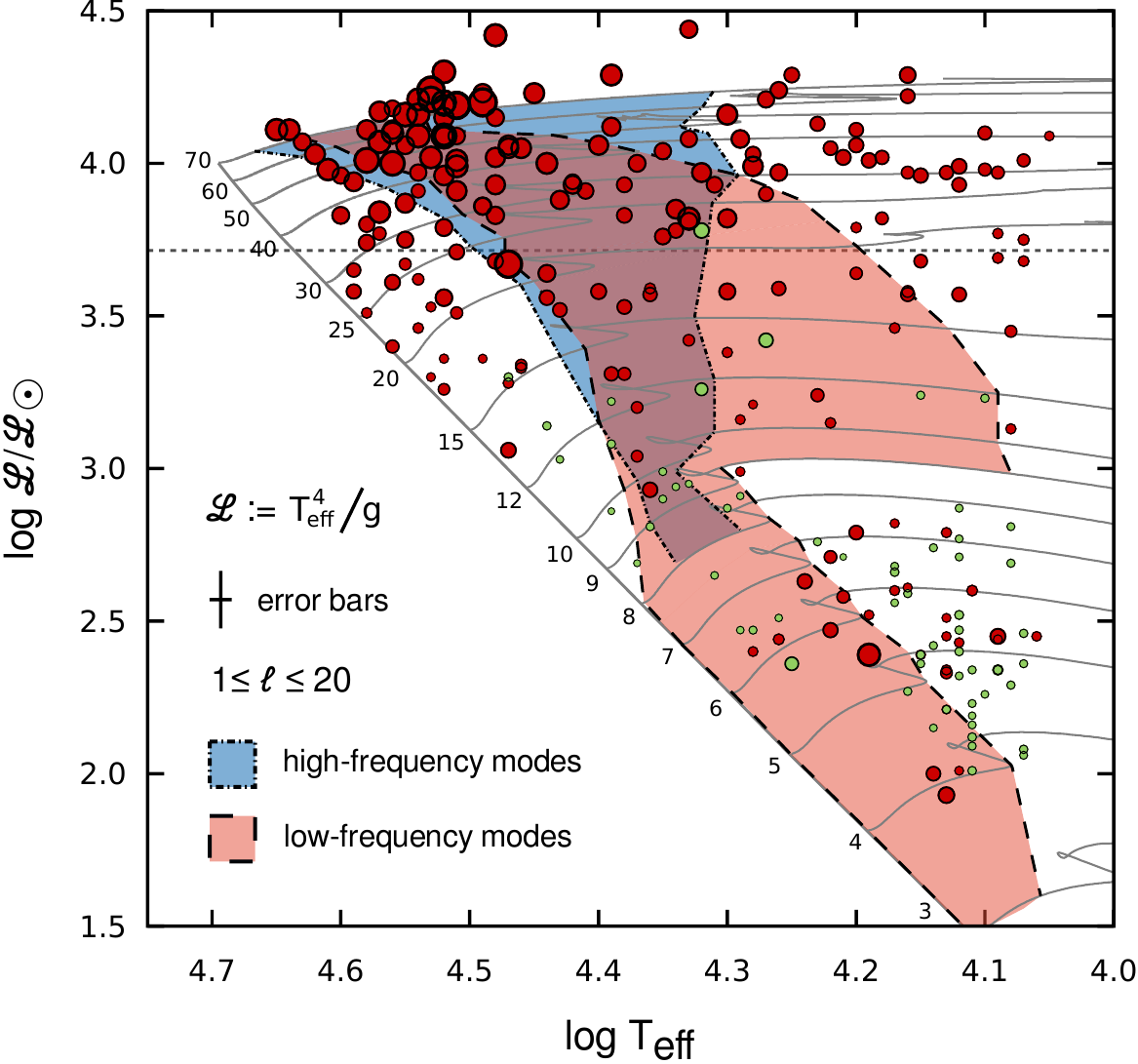}
\caption{
Location in the sHR diagram of the observational dataset presented in \cite{Sim16}: red points stand for the stars presenting an important (or dominant) macroturbulent broadening contribution and green points correspond to stars for which either rotational broadening dominates or both $v$\,sin\,$i$ and $v_ {\rm mac}$ are below 25 km/s \citep[see][for further details]{Sim16}. The size of the points is linearly scaled with $v_ {\rm mac}$, ranging from 10 to 130 km/s. Instability domain predictions (as described in Sects.~\ref{sub_bi_l_equal1} and \ref{sub_bi_l_larger1}) for $1\le\ell\le 20$ are colored in blue and red.
}
\label{fig_inst_lgtab_parameters_stars}
\end{figure}

Hereafter, we work under the hypothesis that the bare minimum requirement to end up in a macroturbulent-type profile produced by pulsation modes is a dense frequency spectrum of excited modes \citep[e.g.][]{Aer09}. In the context of the $\kappa$-mechanism, such spectrum can be achieved by the presence of high-order g modes -- since they are usually excited in an abundant spectrum -- or even by combining the collective effect of a group of modes with a large range of $\ell$ degrees. 
Assuming this requirement, we can determine the best models resulting in such a situation combining the information presented in Fig.~\ref{fig_inst_lgtab_parameters} (instability domains) and Figs.~\ref{fig_models_AI} and \ref{fig_models_JK} (modal diagrams). 
Basically, stars with pulsational properties similar to those of models E -- I plus B and C are good candidates to produce an important broadening of the line profiles (unlike models A and D). 
The former corresponds to regions in Fig.~\ref{fig_inst_lgtab_parameters} colored in red and purple (low-frequency modes) while the latter refers to regions colored in blue (high-frequency modes). 
The region where the existence of an important contribution of the macroturbulent broadening could be explained in terms of high-degree heat-driven modes is, therefore, more extended than when only dipole modes are considered. 
Given the complexity of the interpretation of the results from the pulsational analysis in the coolest post-MS region of the most massive models, we have completely excluded any representation of the instabilities found in that region (see Sect.~\ref{fig_inst_lgtab_parameters}). However, although some excited high-degree modes are found in this region, the associated frequency spectrum is sparse.

Having this in mind, we compare in Fig.~\ref{fig_inst_lgtab_parameters_stars} the empirical distribution of the line-broadening properties of the sample of $\approx$~260 Galactic O and B stars\footnote{From the original sample of $\approx$~430 stars, we exclude the stars for which it was not possible to obtain reliable measurements of the amount of macroturbulent broadening \citep[see notes in][]{Sim16}.} considered by \cite{Sim16} and our instability domain predictions for heat-driven modes with 1 $<\ell\leq$ 20. 
The sample is divided into 2 main groups depending on their global line-broadening characteristics, and the size of the symbols is scaled to $v_ {\rm mac}$, the amount of macroturbulent broadening.
The colors used for the instability domains have the same meaning as in Fig.~\ref{fig_inst_lgtab_parameters}.

As commented in \cite{Sim16}, most of the stars with a remarkable contribution of the macroturbulent broadening (red points) are concentrated in the uppermost part of the sHR diagram and the largest values of $v_ {\rm mac}$ are measured for the late-O/early-B Sgs ($\log \mathscr{L}/\mathscr{L}_{\odot}\approx 4.0$, and $\log T_{\rm eff}\approx$ 4.55\,--\,4.45). 
Many of these stars are located in regions where heat-driven modes could potentially explain the observed line profiles. This refers, in particular, to those targets with $M\,\gtrsim 10\, {\rm M}_{\odot}$ falling inside the instability domain colored in red (corresponding to high-order g modes with a relatively dense global frequency spectrum). 

While the situation is improved when including the full range of mode degrees up to $\ell=$ 20 compared to the case of only considering dipole modes, some problems remain. Indeed, there are still two well populated regions in Fig.~\ref{fig_inst_lgtab_parameters_stars} where stars are clearly located outside the instability domains (even taking individual uncertainties into account) but have line profiles dominated by macroturbulent broadening:
\begin{itemize}
\item the first one refers to the late-B supergiants. We note however that instabilities were found in that region, although they are not directly related to the $\kappa$-mechanism due to the iron opacity bump (see Sect.~\ref{sub_bi_l_larger1}). As a matter of fact, the $\epsilon$-mechanism could play an important role in post-MS models to excite some high-order g modes which can be very well trapped inside the H-burning shell \citep{Scu86,Mor12}. Furthermore, in addition to these modes excited by the $\epsilon$-mechanism, adiabatic and non-adiabatic strange modes are also expected for all masses larger than $\approx 30\,{\rm M}_{\odot}$ (above the dotted line in Fig.~\ref{fig_inst_lgtab_parameters_stars}),  but do not usually lead to a dense frequency spectrum. 
\item the second one refers to the O stars close to the ZAMS with masses $\approx 15$ -- 40~${\rm M}_{\odot}$. This latter point could be used at first glance as a strong argument to dismiss the possibility that heat-driven modes are the main source of macroturbulent broadening in the O star and B supergiant domain. However, before reaching any firm conclusion in this direction, we must explore the effect of considering different assumptions made for the input parameters of the computations on the predicted boundaries of the instability domains. 
\end{itemize}
Finally, we note that there is one region where almost no macroturbulent broadening is found though it is located inside the theoretical instability domain: the low-mass MS stars lying in the SPB instability domain. For this latter group of stars, it is important to notice that, from the observational sample of $\approx$~430 stars presented in \cite{Sim16}, we discarded, in the representation of Fig.~\ref{fig_inst_lgtab_parameters_stars}, stars for which it was only possible to derive a maximum limit on the macroturbulence parameter. When taking these stars into account, we find that most of them are located at the beginning of the MS of  6 -- 9~${\rm M}_{\odot}$ models (where the theoretical computations predict a sparse spectrum of low-order p and g modes), while rather few stars in the theoretical range 4 -- 6~${\rm M}_{\odot}$ were observed (where high-order g modes are expected). 

From these results, it appears that a crucial point for the investigation of the origin of macroturbulence broadening in line profiles resides in the presence (or absence) of 
12 -- 40~${\rm M}_{\odot}$ ZAMS stars with the required pulsational properties. Therefore, we focus in the next section on the efforts needed to excite modes at the beginning of the MS. 

\subsection{Bringing instability domains towards the ZAMS}\label{sec_zams}

\begin{figure}[!t]
\centering
\includegraphics[width=0.49\textwidth]{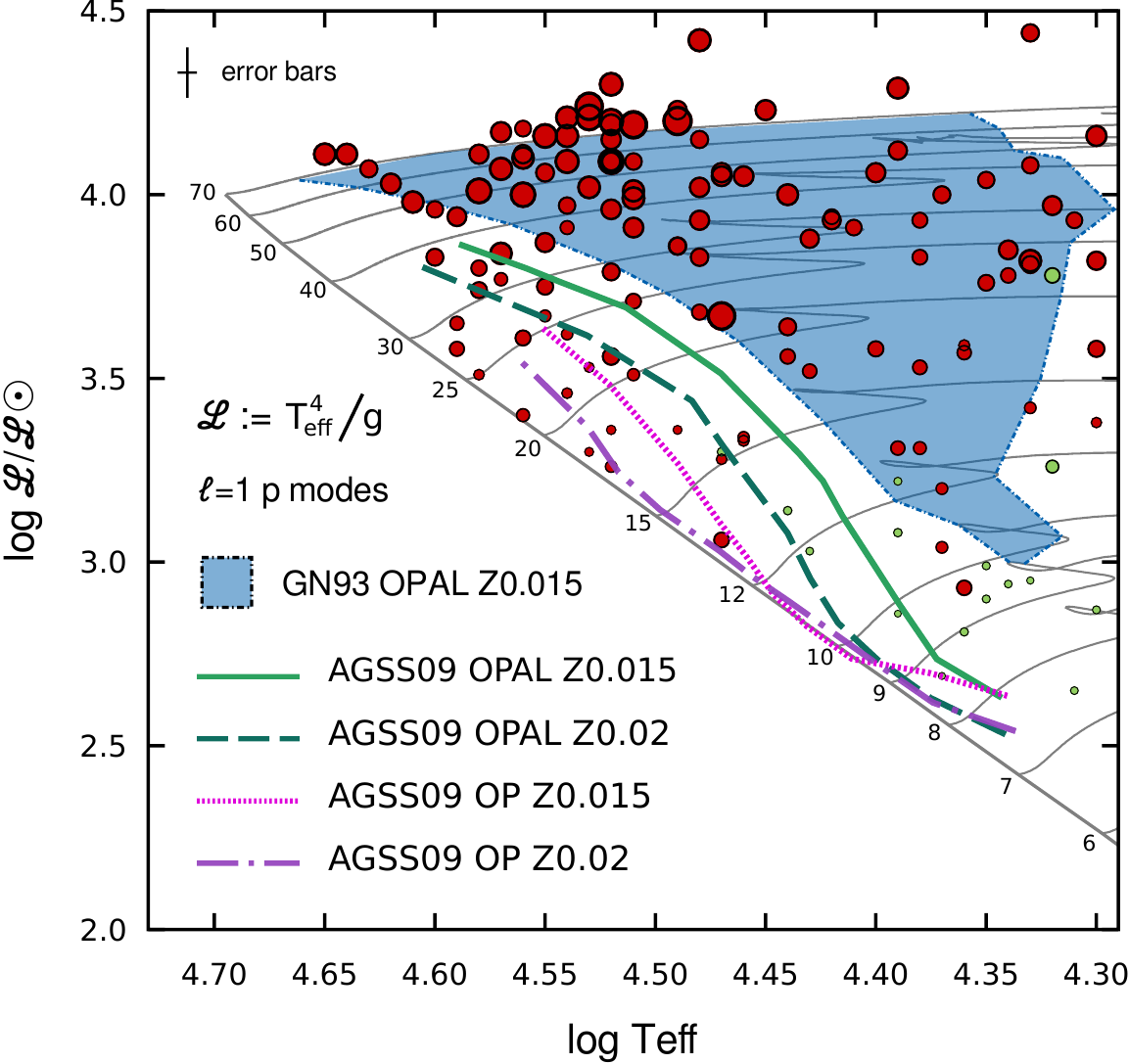}
\caption{Blue boundary of the instability domains for dipole modes. The blue area limited by dotted line represents the instability domains already presented (GN93, $Z=0.015$) while green and purple lines stand for the blue boundaries of the instability domains for models computed with the AGSS09 mixture and different metallicities and opacity tables (OPAL and OP, respectively). Green and red points have the same meaning as in Fig.~\ref{fig_inst_lgtab_parameters_stars} and their sizes are proportional to the macroturbulence velocity.
}
\label{fig_cles}
\end{figure}

As just concluded, any scenario aiming at producing a macroturbulent-type profile in 12 -- 40~${\rm M}_{\odot}$ stars must be efficient in their early evolution.
As illustrated in the literature, the use of a different set of input parameters (e.g. overshooting parameter, opacities, metal mixture, metallicity) or even evolutionary codes may alter the results of our computations \citep[see, e.g.,][for recent illustrations of these effects]{Miglio2007a, Miglio2007b, Pamyatnykh2007,Zdravkov2008, Mar13, Cas14, Mor16}.
The question is thus the following: how can we widen the instability strip towards the ZAMS? 
While the overshooting parameter will be of no help to stretch the blue boundary, the metallicity, the detailed chemical mixture (e.g. \citeauthor[][]{Asplund2009} \citeyear[][]{Asplund2009}, hereafter AGSS09 vs. \citeauthor[][]{Grevesse1993} \citeyear[][]{Grevesse1993}, hereafter GN93) and opacities are obviously playing a major role in the efficiency of the $\kappa$-mechanism \citep{Miglio2007a,Pamyatnykh2007,Mor16}.

The computations presented in this paper were performed with the GN93 metal mixture and adopting a metallicity of 0.015, in agreement with AGSS09. The choice of this combination was motivated by the fact that the AGSS09 solar mixture  was not yet implemented in the ATON evolutionary code. 
In order to investigate the effect of the use of a different metal mixture (AGSS09 vs. GN93), metallicity (0.015 vs. 0.020) and opacities (OP vs. OPAL) we started the computation of new models in the 7 to 40~${\rm M}_{\odot}$ range combining the 
CLES evolution code from Li\`ege \citep[][for which both opacity tables and both metal mixtures are implemented]{Scu08b} and the non-adiabatic code MAD.
The models computed with OP opacities are limited to masses up to 25~${\rm M}_{\odot}$ due to the narrower temperature and density range in the OP tables. 
Except for the three set of parameters indicated above, the rest of input physics were fixed to the ones used for our ATON models. While the presentation of these new (still on going) computations will be subject of a more extensive study in the future, we focus here for illustrative purposes on the predictions for the blue boundary of the p-mode ($\omega$=1--25), $\ell$=1 instability domain. 

Figure~\ref{fig_cles} summarizes the results of these new computations. The instability domain derived with the ATON models (discussed in Sect.~\ref{sec_instabilitystrips}) is represented by the area colored in blue, while the blue boundaries obtained with CLES+MAD under various assumptions are indicated with the different lines. This figure illustrates how critical is the selection of the three input parameters mentioned above on reaching a definitive conclusion about the heat-driven mode scenario. First, the comparison between the ATON and CLES computations with $Z=0.015$ and OPAL allows to illustrate the effect of the metal mixture. Although no modes are found close enough to the ZAMS yet, the AGSS09 boundary is clearly shifted towards higher $T_{\rm eff}$ (compared to the GN93 one), and the associated instability strip now gathers some more O stars presenting a dominant macroturbulent broadening contribution. This result is encouraging and brings us one step further. Indeed, the situation improves even more (for a fixed metallicity) when considering the OP opacities (as already pointed out by \citealt{Miglio2007a} and \citealt{Pamyatnykh2007}, the combination AGSS09+OP favors the excitation through the $\kappa$-mechanism due to the iron opacity bump much more than the combination GN93+OPAL). Last, the effect of metallicity is clear: the larger the metallicity the broader the instability domain.

These effects (increasing metallicity and opacity) both broaden the instability region in our calculations and thus help both supporting the pulsational hypothesis as the origin of the macroturbulent broadening. However these results must be taken with caution. 
First, given the most recent value of metallicity ($Z=0.014$) derived from the spectroscopic analysis of a sample of $\approx$~30 early B-type stars in the solar vicinity by \citet[][$Z$=0.014]{Nie12}, the higher metallicity ($Z=0.02$) favoring a wider instability domain may not be representative of massive stars. 
Secondly, concerning the opacities, we also consider of interest mentioning the recent studies of \citet{Salmon2012} and \citet{TurckChieze2013}, who suggested the need to reassess opacity computations, in particular for nickel, that showed to affect importantly the $\kappa$-mechanism instability predictions in massive stars \citep[see also][who revealed a larger mean Rosseland opacity in their OPAS computations of 11\% -- compared to OPAL and OP -- around $\log T\simeq 6.2$ in solar modeling]{Mon15}.
Moreover, the recent work of \citet{Wal15} has shown that the newly developed OPLIB opacity tables from Los Alamos National Laboratory induce an even higher Rosseland mean opacity coefficient at the maximum of the iron opacity bump (see Figs.~1 and 3 of their paper). 
Lastly, \citet[][]{Bailey2015} found a large disagreement between the experimental measurement of iron opacity and theoretical computations, for conditions close to that of the base of the solar convective zone. Based on this, \citet{Mor16} showed that a 75\% increase in Fe and Ni monochromatic opacities  from OP tables extends the $\beta$ Cephei domain to the ZAMS, in agreement with our results without any metal enhancement presented in Fig. \ref{fig_cles}.
As a result, while the metallicity cannot play a major role in widening the instability strip since the value fixed in the computations already correspond to the commonly accepted metallicity value for massive stars, the computations of the opacities constitute therefore the crucial point to help widen the instability domains towards the ZAMS. 

Concluding, our investigation on the instability domains associated with high-degree modes, in combination with the results presented in Fig.~\ref{fig_cles} and all the studies on the opacity quoted above show that the heat-driven mode scenario to explain macroturbulent broadening in O and B stars may not be in a situation as bad as when only considering the dipole modes. Of particular importance is the fact that certain assumptions on the opacities can widen the instability domains towards the ZAMS, allowing some of the O stars with macroturbulent broadening dominated profiles to lie inside.
However, there are still some clues which make us suspect that the heat-driven pulsation modes due to the $\kappa$-mechanism in the iron bump cannot explain alone the distribution of line-broadening properties of the line profiles in massive stars
First, in view of panels A and D in Fig. \ref{fig_models_AI}, even assuming the opacities resulting from the most favorable conditions, it appears that a dense spectrum of pulsation modes is very difficult to be produced at the beginning of the MS. Second, even if we assume that models similar to those labeled with B and E could be found close enough to the ZAMS, we would need to explain why there is a non-negligible number of narrow line (i.e. with a negligible macroturbulent broadening contribution) MS stars with masses in the range $\approx$~9 -- 20~${\rm M}_{\odot}$. 

We briefly discuss in the next section some additional hypotheses to attempt to explain the origin of the non-rotational extra-broadening in OB stars.

\subsection{Which additional driving scenario?}\label{sec_sources}

In addition to the spectroscopic signature of the macroturbulent-type profile, direct observations also indicate the presence of pulsations in stars outside or close to the blue instability boundary for masses between 15 -- 40~${\rm M}_{\odot}$. Indeed, the presence of line-profile variations has been demonstrated to be a common feature among O-type stars, even close to the ZAMS \citep[e.g.][]{Ful96}. For some of them it has been possible to link this variability with pulsations \citep[e.g.][]{Blomme2011,Briquet2011,Mahy2011}. Moreover,
there is increasing observational \citep{Degroote2010} and theoretical evidence \citep{Belkacem2010,Samadi2010} for the presence of stochastic oscillations. Furthermore, strange modes \citep[e.g.][]{Noe08,Gla09,Saio2009}, oscillatory convective modes \citep{Saio2011}, and pulsation modes due to the $\epsilon$-mechanism \citep{Scu86}, have been suggested to be the explanation of photometric observations \citep{Aer10,Mor12}. All these mechanisms may help producing a macroturbulent-type profile, though they usually produce a scarce spectrum of unstable modes. While several other scenarios have been suggested for the stellar oscillation origin, the most viable to this date is the stochastically driven waves induced by the turbulent pressure fluctuations in the interface between the convective core and the envelope \citep{Aer15} or in the sub-surface convective zone \citep{Gra15}. 
In particular, \citet{Aer15} demonstrated with CoRoT space photometry the observational evidence for the occurrence of convectively driven internal gravity waves in young massive O-type stars giving rise to a macroturbulent-type line profile. \citet{Gra15} have computed and mapped the turbulent pressure associated to sub-surface convection zones in massive MS and post-MS models. The authors presented a good correlation between $P^{\rm max}_{\rm turb}/P$ and $v_ {\rm mac}$. 
As indicated by \citet{Sim16}, the interplay between the turbulent pressure fluctuation hypothesis and the heat-driven modes due to the $\kappa$-mechanism scenario seem like a promising explanation to the origin of macroturbulent broadening in the upper part of the sHR diagram.

\section{Summary and concluding remarks}\label{sec_conclusions}
In this paper we investigate the pulsation modes driven by the $\kappa$-mechanism in the iron opacity bump in massive stars. 
One of the goals of this paper is to provide new predictions for the instability domains of non-radial modes driven by the $\kappa$-mechanism in the opacity bump of iron. The originality of our results comes from the use of a large domain of $\ell$  degrees for a large range of masses between 3 and 70~${\rm M}_{\odot}$.
Altough these modes are not visible with photometric detections, we think our results will be a valuable tool for constraining the new high-quality spectroscopic observations in massive OB stars, as e.g. the observational material gathered by the IACOB project. In particular, we provide a global overview of the frequency ranges and characteristics of the modes to expect on the main sequence and the post-main sequence of these massive models.  
On the main-sequence, a few high-frequency modes are found unstable in our theoretical models with periods around $8$ hours. These periods increase during the evolution from the main sequence to the post-main sequence, but the frequency spectrum is always characterized by a small number of degree $\ell$ modes, with $\ell$ never larger than 10. On the post-main sequence (and even on the main-sequence for the smallest masses), low-frequency modes are excited in a much more abundant spectrum. These modes have periods ranging between a couple of hours to a dozen of days. We find that the presence of the low-frequency modes in the evolved models which do not present any intermediate convective zone (appearing for models $\ge 12\,{\rm M}_{\odot}$) is due to the sharp feature already present in the Brunt-V\"{a}is\"{a}l\"{a} frequency in models of 9~${\rm M}_{\odot}$, in agreement with \citet{Das13}. 
We discuss the appearance of these low-frequency unstable modes by considering the constraints on the efficiency of the $\kappa$-mechanism. Actually, we find that for increasing $\ell$, the range of unstable low-frequency modes moves towards higher frequencies. Furthemore, by increasing the degree of the modes, the blue boundary of the instability domains for the low-frequency modes is shifted towards the zero age main sequence, even overlapping the high-frequency instability domain ($\beta$ Cephei instability strip). 

In addition to the usual spectroscopic observations, the emergence in the last few years of the investigation of the so-called macroturbulent broadening has provided a new way, less expensive in terms of observational time,  to study stellar oscillations. While the strongest hypothesis for the origin of this macroturbulent-type profile is to relate it to stellar oscillations \citep[e.g.][]{Sim16}, several mechanisms and types of waves/modes driving the macroturbulence have been suggested. 
In this paper we show that pulsation modes driven by the $\kappa$-mechanism in the opacity bump of iron cannot explain alone the distribution of line-broadening properties of the line profiles in the whole OB star domain, and hence they might not be the dominant agent producing the so-called macroturbulent broadening in O stars and B Supergiants.
Indeed, although we consider numerous spectra of modes combining different $\ell$ degrees, the location of the instability domains for this kind of spectra do not entirely coincide with the location  of the high $v_ {\rm mac}$ stars. 
We investigate how the physics of the models have to be adapted in order to provide such an excitation in the beginning of the main sequence, as we think it is the key point for constraining the origin of the mechanism broadening the line profiles. While the major outcome would be achieved considering new computations of the opacities, it is still difficult to obtain such an abundant spectrum close to the zero age main sequence. Indeed, for a reasonable metallicity in massive stars ($Z\sim 0.015$), models close to the beginning of the main sequence are found unstable when considering the OP opacities, although the frequency spectrum of these models is scarce.   
Another additional mechanism should therefore trigger the broadening of the line profiles and it must be efficient in the early evolution of the star, close to the beginning of the main sequence \citep[such as the scenarios of][]{Gra15,Aer15}.  
Already foreseen by \citet{Sim16}, we agree therefore with the idea that the line broadening must be due to a combination of different sources depending on the regions in the HR diagram. 

We propose several directions for future investigations: (a) the possibility of extending the instability boundaries for stars with $M \ge 12\,{\rm M}_{\odot}$ towards the zero age main sequence with a reasonable enhancement of opacities (especially nickel) though the calculations computed until here have shown that few modes are excited in these regions, 
(b) the study of other driving mechanisms for stellar oscillations not accounted for in the present computations, as e.g. theoretical simulations to confirm that the turbulent pressure fluctuations investigated by \citet[][]{Gra15} do indeed end up in a macroturbulent profile: in particular, if an excitation mechanism such as pressure fluctuations (which in 1-D models  correspond to a radial velocity pattern) can give rise to macroturbulent velocity distributions as used here with a strong tangential component and (c) the extension of the type of comparison between observations and models presented in this paper but including information about line-profile variability as extracted from adequate time-resolved spectroscopic observations \citep[see, e.g., first results in this direction in][]{Sim16}.

\begin{acknowledgements}
This work has been funded by the Spanish Ministry of Economy and
  Competitiveness (MINECO) under the grants AYA2010-21697-C05-04, AYA2012-39364-C02-01, 
  and Severo Ochoa SEV-2011-0187, and by the Research Council of KU\,Leuven under grant GOA/2013/012. We thanks C. Aerts, P. Degroote, J. Montalban, E. Moravveji, J. Puls and J. Telting for interesting and fruitful discussions related to the work. 
\end{acknowledgements}




%
\end{document}